\documentclass[%
aip,
amsmath,amssymb,
reprint,%
]{revtex4-1}

\usepackage{graphicx}
\usepackage{dcolumn}
\usepackage{bm}

\usepackage[utf8]{inputenc}
\usepackage[T1]{fontenc}
\usepackage{mathptmx}
\usepackage{etoolbox}
\usepackage{siunitx}
\usepackage{graphicx}
\usepackage{float}
\usepackage {CJK}

\makeatletter
\def\@email#1#2{%
	\endgroup
	\patchcmd{\titleblock@produce}
	{\frontmatter@RRAPformat}
	{\frontmatter@RRAPformat{\produce@RRAP{*#1\href{mailto:#2}{#2}}}\frontmatter@RRAPformat}
	{}{}
}%
\makeatother

\newcommand{\srate}{\dot{\gamma}}
\newcommand{\su}{\si{s^{-1}}}
\newcommand{\Dt}{\Delta t}
\begin{document}
	
\begin {CJK*} {UTF8}{gbsn}	
	\preprint{AIP/123-QED}
	
	\title{Boundary conditions alter density and stress fluctuations in shear-thickening suspensions}
	\author{Meng-Fei Hu（胡梦菲）}
	\affiliation{State Key Laboratory for Strength and Vibration of Mechanical Structures,\\ School of Aerospace Engineering, Xi'an Jiaotong University, Xi'an 710049, China
	}%
	
	\author{Song-Chuan Zhao（赵松川）}%
	\email{songchuan.zhao@outlook.com}
	
	\altaffiliation{Author to whom correspondence should be addressed}
	\affiliation{State Key Laboratory for Strength and Vibration of Mechanical Structures,\\ School of Aerospace Engineering, Xi'an Jiaotong University, Xi'an 710049, China
	}%
	
	\date{\today}
	
	\begin{abstract}
		Discontinuous shear thickening (DST) in dense suspensions is accompanied by significant fluctuations in stress at a fixed shear rate. In this work, normal stress fluctuations are shown to have a one-to-one relationship with the formation and dissolution of local high-density regions. Namely, a burst in the force response corresponds to the spontaneous appearance of inhomogeneity. We observe that boundary conditions can significantly alter the spatiotemporal scale of these fluctuations, from short-lived to more sustained and enduring patterns. We estimate the occurrence frequency $R$ and the average intensity $Q$ of individual bursts/inhomogeneity events. The growth of $R$ with the shear rate is the most rapid for the rigid boundary, whereas $Q$ is nonmonotonic with confinement stiffness. Our results indicate that boundary conditions alter the development of inhomogeneity and thus the stress response under shear.
	\end{abstract}
	
	\maketitle
\end{CJK*}

	\section{\label{sec:level1}Introduction}
	
	Particulate suspensions, as multiphase systems, are ubiquitously found in both natural and engineering contexts~\cite{Pistone2015, Guergen2017}. The flow behavior of these dense suspensions is notably complex, characterized by a prominent non-Newtonian feature known as Discontinuous Shear Thickening (DST)~\cite{Brown2014,Wyart2014,Seto2013,Peters2016}. The shear thickening phenomenon in non-Brownian suspensions is attributed to a transition from a frictionless to a frictional flow state~\cite{Clavaud2017,Wyart2014,Seto2013,Mari2014}. Simulations have demonstrated that when the critical stress threshold is surpassed, frictional contact networks form within the suspension~\cite{Sedes2020,Rahbari2021,Goyal2024,Naald2023,Nabizadeh2022}. It is suggested that a hallmark of these frictional clusters is the emergence of a positive first normal stress difference under shear in experiments~\cite{Lootens2005,Brown2012,Xu2020,Andrade2020}. Boundary confinement is required to counterbalance these stresses; otherwise, the particle phase undergoes dilation, leading to the collapse of the contact network and the subsequent elimination of DST~\cite{Brown2012}. Boundary confinement is, therefore, considered crucial for the occurrence of DST~\cite{Maharjan2021,Brown2014,Fall2012}.

	Another feature accompanying DST is significant temporal fluctuations in stress or shear rate~\cite{Lootens2003,Hermes2016,Xu2020,Sedes2020}. These global fluctuations initially suggest a temporal transition between two flow states~\cite{Lootens2003}. However, recent experimental and simulation advancements propose a different scenario. In a system experiencing DST, the uniform state is inherently unstable, leading to the emergence of inhomogeneities. These inhomogeneities may manifest as stress waves~\cite{SaintMichel2018,Chacko2018,Gauthier2023}, density waves~\cite{Ovarlez2020,Shi2023,Bougouin2024}, torque clusters~\cite{Rahbari2021}, or localized jammed regions~\cite{Rathee2017,Rathee2020,Hu2022,Rathee2022,Shi2024}. Their evolution underlies the measured fluctuations. Given the critical role of boundary confinement in DST, it is plausible that boundary conditions could influence the development of inhomogeneities~\cite{Shi2023,Shi2024}.


	The present study investigates the temporal measurements of normal force responses in a cornstarch suspension subjected to shear under varying boundary constraints. The force measurement is accompanied by simultaneous observations of inhomogeneity. The primary emphasis of our research is to understand the impact of boundary flexibility on the response of normal stresses. The key findings include: (1) The bursts in normal force response are associated with instantaneous density inhomogeneities beyond the onset of DST. (2) Flexible boundaries promote an extended duration and, in certain instances, amplified intensity of bursts. (3) The evaluation of the normal force response is conducted under various boundary stiffness conditions by assessing the average duration, occurrence rate, and intensity of bursts.

	\section{experimental protocol}
	\label{sec:exp}
\begin{figure}
		\centering
		\includegraphics[scale=1]{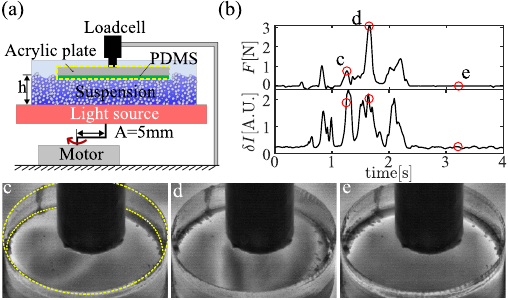}
		\caption{(a) A schematic of the experimental setup. (b) The time-series diagrams of normal force, $F$, and light intensity variation, $\delta I$, under the hard PDMS boundary at $\dot \gamma=24.3 \si{s^{-1}}$. Here, $\delta I$ of the transmission light measures inhomogeneity in the suspension. (c-e) Example of snapshots where $\delta I$ are calculated from. The dark areas indicate high-density regions. Dotted lines mark the edge of the top boundary. (Multimedia view in Movie S1) }
		\label{fig:figure1}
	\end{figure}

	The experimental setup comprised a dense suspension of cornstarch particles (Aladdin S116030) immersed in a density-matched solution composed of deionized water and cesium chloride. The volume fraction $\phi$ is fixed at $43\%$. The starch particles are polydispersed, with an average diameter $d\approx 15~\si{\mu m}$. The suspension is contained within a cylindrical vessel of a diameter of $120~\si{mm}$. The container is subject to an orbital motion with a radius $A=5\,\si{mm}$ (Fig.~\ref{fig:figure1}a). The motion is primarily horizontal, with a vertical component of less than 4\%. A top shear plate of diameter $D$ is mounted within the laboratory reference frame. The gap size between the top plate and the bottom, denoted as $h$, is modulated within the range of $3.5-5~\si{mm}$, more than two orders of magnitude larger than $d$. The shear rate $\dot{\gamma}=2\pi f A/h$ is regulated by adjusting $h$ and the frequency of orbital motion $f$, ranging from $0.5$ to $8.33~\si{Hz}$. This shearing methodology is distinct from conventional parallel-plate rheometry. As the orbital motion of the bottom plate is purely translational, no inherent gradient prompts inhomogeneities. 	
	
	A load cell (Honeywell 31) records the normal force applied to the upper boundary at a sampling frequency of $1000~\si{Hz}$. The cornstarch suspension is illuminated from below with monochrome light. A high-speed camera (Microtron EoSens 1.1cxp2) synchronously captures the transmitted light along with the force measurement. As demonstrated in Ref.~\citenum{Shi2023}, the light transmittance provides a measure of local particle density. This technique is utilized in this study to qualitatively assess the inhomogeneity in the suspension. Figure~\ref{fig:figure1}a depicts the experimental protocol.
	
	To examine the influence of boundary conditions, a bare acrylic plate (rigid) and optional layers, including Polydimethylsiloxane (PDMS) films and a silicone oil layer of 100~\si{cSt}, are used. All additional layers have a thickness of 2~\si{mm}. The PDMS layer is prepared by mixing the precursor polymer and curing agent (Sylgard 184, Dow corning). The mixing ratio and the curing process controls the stiffness of the PDMS layer~\cite{Trappmann2012}. This study employs PDMS of two hardness levels, namely hard PDMS and soft PDMS. The hard PDMS layer was prepared by a 35:1 weight ratio and baking at 80\si{\celsius} for 15 minutes. Soft PDMS was produced at a weight ratio of 85:1 and cured at 70 \si{\celsius} for 14 hours. The  Young's moduli of these two PDMS layers, $E_h=11~\si{kPa}$ and $E_s=1.3~\si{kPa}$, are measured with oscillatory rheometry. 
	For experiments with the oil layer, the shear rate, $\srate$, accounts for the thickness of the additional layer, which also experiences shear flow. However, due to the rate-dependent viscosity of the suspension, $ \srate$ is underestimated at relatively high shears.
	
The open lateral boundary of the shear zone in our protocol allows the influence of external flow under the free surface, where the shear profile is nonlinear~\cite{Shi2023} distinct from the flow regulated by the top plate. The magnitude of this influence is expected to scale with the periphery length, $\pi D$. The force response is confirmed to be extensive with the area of the top plate, $\pi D^2/4$, rather than the periphery length (refer to Appendix~\ref{app:bsize}). Moreover, the average force response of our protocol aligns with that measured in the rheometer (data in Appendix~\ref{app:rheo}), justifying our experimental configuration. In consequence, edge effects are disregarded. Another potential consequence of the open boundary is the particle migration out of the shear zone at high shears, which will be addressed with the force measurements in Section~\ref{sec:rigid}. The findings presented herein correspond to a fixed top plate diameter $D=20~\si{mm}$.

	\section{Normal Force Measurements}
	
	Figure~\ref{fig:figure1}b-e illustrates a temporal profile of the normal force, $F(t)$, on the hard PDMS confinement and the snapshots of the suspension at corresponding moments. The grayscale levels in Figure~\ref{fig:figure1}c-e act as indicators of local particle density; that is, the darker the region, the denser it is. Transient high density regions randomly emerge. These regions are unstable and undergo a cycle of formation, evolution, and eventual dissolution. These cycles are referred to as density inhomogeneity events. The variance of light intensity in the shear zone, $\delta I$, which indicates the degree of inhomogeneity, is plotted against time in Fig.~\ref{fig:figure1}b.  
	It is remarkable that the emergence of high-density regions (peaks in $\delta I$) coincides with significant increases in the amplitude of $F$ (Movie S1). As these dense regions disperse, the signal reverts to its baseline level. The correspondence between spontaneous inhomogeneity and the finite normal force is consistent throughout the entire profile. Since the normal stress results from percolating particle clusters, this correlation represents a potential relationship between macroscopic force fluctuation and the arrangement of microstructures within the suspension. We proceed to examine the characteristics of $F(t)$ under varying boundary conditions, starting with the rigid boundary provided by the acrylic plate.

	\subsection{Rigid boundary}
	\label{sec:rigid}
	
	\begin{figure}
		\centering
		\includegraphics[scale=1]{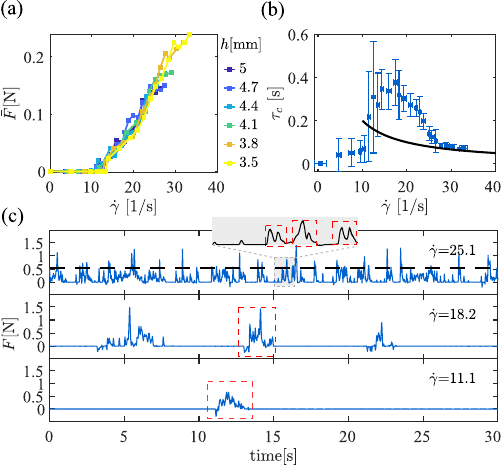}
		\caption{Rigid boundary. (a) Time average of $F(t)$  as a function of shear rate $\dot\gamma$ for different $h$. (b) Evaluation of autocorrelation time scales $\tau_c$ as shear rate $\dot\gamma$. Error bars correspond to the standard deviation taken over all depths. The black line denotes $2/{\dot\gamma}$. (c) Time series of $F(t)$ for several imposed shear rates $\dot\gamma$. The dashed line represents the 95th percentile of $F$. The typical elementary impulses identified by the approach in Sec.~\ref{sec:element} are denoted by dashed red boxes. }
		\label{fig:rigid_overall}
	\end{figure}

	Figure~\ref{fig:rigid_overall}a illustrates the time-averaged normal force response, $\bar{F}=\langle F(t)\rangle$, for various shear rates, $\dot\gamma$, and gap heights, $h$. A key observation is the consistent collapse of $\bar{F}$ across different $h$ values at a given shear rate $\dot{\gamma}$. This suggests that $\dot{\gamma}$ is the primary control parameter. 
	At lower shear rates, the force profile $F(t)$ exhibits steady white noise around $\bar{F}=0$. A critical shear rate, $\dot\gamma_c=11.4 \pm 1.2\si{s^{-1}}$, marks the onset of a finite normal force response. This critical value aligns with the DST transition, as independently verified using parallel-plate rheometry (refer to Appendix~\ref{app:rheo}). Beyond this critical shear rate, $\bar{F}$ exhibits an approximately linear increase with $\dot\gamma$. However, a decrease in the response is observed beyond a certain shear rate, which is associated with $h$ (full data set available in Appendix~\ref{app:rheo}). We attribute this decrease at high shears to the migration of particles from the high-stress zone under the upper plate towards the region under the free surface. High-shear rate data exhibiting this dilated subset will be excluded from subsequent analyses.

	The increase of $\bar{F}$ with $\srate$ is dictated by the variations in the force profile $F(t)$. Similar to that depicted in Fig.~\ref{fig:figure1}, the rise of $F$ coincides with the emergence of density inhomogeneity. Under the rigid boundary, however, the density variation is less intense, occurs more rapidly, and manifests on a smaller spatial scale (refer to Supplemental Movie S2). The force profile is characterized by intermittent impulse bursts against a zero background. At $\dot\gamma=11.1 \su$, slightly above the critical shear rate $\dot\gamma_c$, distinct intervals of high stress become apparent (see Fig.~\ref{fig:rigid_overall}c). As $\srate$ increases, so does the frequency of these bursts, which is consistent with the experimental observations of shear stress response in Ref.~\citenum{Rathee2020}. Furthermore, the majority of these bursts transition from broad envelopes at low $\srate$ to individual pulses at high $\srate$, as illustrated in Fig.~\ref{fig:rigid_overall}c.

	We calculate a timescale, $\tau_c$, via the short-time decay of the autocorrelation of $F(t)$. As the white noise background does not contribute to the autocorrelation, $\tau_c$ quantifies the duration of impulse bursts that occur randomly. As depicted in Fig.~\ref{fig:rigid_overall}b, $\tau_c$ undergoes a jump at $\srate_c$, reaching its maximum of approximately $5/{\srate_c}$ near $\srate=\SI{17}{s^{-1}}$ before declining. The pattern of $\tau_c$ aligns with the variation of $F(t)$ shown in Fig.~\ref{fig:rigid_overall}c. In particular, $\tau_c$ converges towards ${2}/{{\dot\gamma}}$ beyond $\srate\approx 25\su$. The width of individual peaks also follows the same scaling of approximately $1/\srate$,~\cite{Lootens2003} indicating the dominance of individual pulses in the normal force fluctuations under high shears, as illustrated in Fig.~\ref{fig:rigid_overall}c. These pulses are asymmetric peaks, characterized by a more abrupt rise than decline, as detailed in Appendix~\ref{app:subsec}.
	
	Fluctuations in $F$ indicate the percolation and disruption of the frictional force network. Under conditions of high $\srate$, the shear timescale, $1/\srate$, governs the rearrangement dynamics of force networks. In contrast, a longer $\tau_c$ in low $\srate$ conditions implies a substantial correlation in the rearrangement dynamics, suggesting that a rearrangement in one location, underlying a pulse, is likely to trigger another elsewhere. In other words, a subsequent pulse likely occurs before the initial pulse completely diminishes. This scenario is demonstrated by pulses of width similar to $1/\srate$ that overlay the impulse envelope, as shown in Fig.~\ref{fig:rigid_overall}c. The observed trend of $\tau_c$ is consistent with the spatial correlations as recently reported~\cite{Xu2020,Nabizadeh2022,Wang2020}. The length scale of these correlations, which reflect the sizes of the frictional contact clusters, shows a notable increase as approaching the shear-thickening regime, followed by a subsequent decrease.

	\subsection{Flexible boundaries}
	
\begin{figure}
		\centering
		\includegraphics[scale=1]{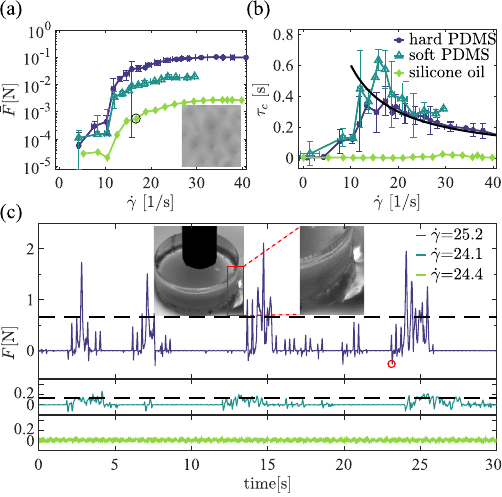}
		\caption{Flexible boundaries: light green diamonds $\diamond$, dark green triangles $\triangle$ and purple circles $\circ$ represent silicone oil, soft PDMS, and hard PDMS boundaries respectively. (a) Time-average force $\bar{F}$ versus the shear rate $\srate$. The inset shows the persistent inhomogeneous state under the silicone oil layer at the shear rate marked by the black circle. (b) Autocorrelation time scales $\tau_c$ for all flexible boundaries. The black line is ${6}/{{\dot\gamma}}$. Data of different $h$ are averaged in (a) and (b), and error bars indicate the standard deviation. (c) Time series of $F(t)$ for the three boundaries at $\srate\approx \SI{25}{\second^{-1}}$. The inset corresponds to the moment of a negative signal, highlighted by the red circle. The dashed line represents the 95th percentile of $F$.}
		\label{fig:soft_overall}
	\end{figure}

	Figure~\ref{fig:soft_overall} illustrates the response of the normal force under flexible boundaries. For all types of soft boundaries, the average force, denoted as $\bar{F}$, increases at the same critical shear rate, $\srate_c$, as under the rigid boundary. This suggests that the formation of frictional clusters is an intrinsic characteristic of the suspension, and the onset of shear-thickening does not depend on the boundary conditions.
	However, it is important to note that the boundary flexibility does influence the magnitude of $\bar{F}$. As the boundary becomes softer, the growth of $\bar{F}$ with the shear rate ($\srate$) becomes weaker. Most notably, the type of boundary has a significant impact on the development of inhomogeneity and the fluctuations in $F(t)$.

	The oil boundary presents a particularly outstanding case, as neither the stress nor the flow velocity at the interface remains fixed. Due to the potential underestimation of $\srate$ (as explained in Sec.~\ref{sec:exp}), we emphasize its qualitative characteristics at this stage. Here, we observed a sustained non-uniform distribution of density for $\srate > \srate_c$, as depicted in Fig.~\ref{fig:soft_overall}a inset. This nonuniformity manifests as localized density waves that propagate along circular trajectories and self-organize into a hexagonal pattern~\cite{Shi2023,Shi2024}. Despite the significant density heterogeneity, the normal force does not exhibit any bursts or pulses. Instead, the force $F(t)$ fluctuates consistently around a finite $\bar{F}$ value, as depicted in Fig.~\ref{fig:soft_overall}c. The featureless fluctuations result in a correlation time $\tau_c$ that is indistinguishable from zero across the entire range of $\srate$ (refer to Fig.~\ref{fig:soft_overall}b). This unique state further supports the correlation between impulse bursts in $F(t)$ and the transitory of density heterogeneities, indicating that the absence of one implies the absence of the other.
	
	The suspension beneath the oil layer is constrained by the interfacial tension, denoted as $\Gamma=27.1~\si{mN/m}$. The interface deforms to counterbalance the normal stresses accompanying inhomogeneity. According to Ref.~\citenum{Shi2023}, the formation of density waves can be attributed to a balance between particle aggregation, which is a result of local viscosity proliferation, and the outward migration from regions of high stress. The characteristic stress here, represented as $\sigma^*\approx 9\si{Pa}$, corresponds to the onset stress of DST (Appendix~\ref{app:rheo}). The total normal force is anticipated to be $\bar{F} \sim \sigma^*\ a\ n$, where $a$ and $n$ denote the size and number of individual density waves, respectively. Our study confirms that the normal stress, calculated as $\sigma = \bar{F}/a\ n\approx \sigma^*$, matches this anticipation. Furthermore, it has been shown that the wavenumber of the hexagonal pattern, and consequently $n$, increases with $\srate$,~\cite{Shi2023} which primarily contributes to the growth of $\bar{F}$ beyond $\srate_c$ for the oil boundary.

	In Fig.~\ref{fig:soft_overall}a-b, the average normal force $\bar{F}$ and the autocorrelation time $\tau_c$ for PDMS boundaries qualitatively reproduce the trend observed with the rigid plate in Fig.~\ref{fig:figure1}. However, PDMS boundaries demonstrate lower $\bar{F}$ and longer $\tau_c$ values. The longer $\tau_c$ suggests a stronger correlation of the inhomogeneity. As observed in Supplemental Movie S1, high-density regions span wider areas and take longer to dissipate, leading to longer-lasting bursts than the rigid boundary. Considering these prolonged bursts, it may be intuitively concluded that the flexibility of the boundaries causes a reduction in the strength of pulses in $F(t)$ accounting for the lower $\bar{F}$. In Fig.~\ref{fig:rigid_overall}c and Fig.~\ref{fig:soft_overall}c, we defined the 95th percentile of the normal force response, $F_{high}$, as a pulse strength indicator for all non-fluid boundaries. Interestingly, the hard PDMS layer produced the highest $F_{high}$, even exceeding that of the rigid plate. This finding suggests that the dependence of $\bar{F}$ and pulse strength on boundary stiffness do not align. 
	
	Furthermore, the hard PDMS boundary results in more noticeable negative pulses, as depicted in Fig.~\ref{fig:soft_overall}c. These negative signals, which always follow intense positive pulses, result from the fracture of locally jammed suspension near the edge of the top plate~\cite{Roche2013,Gauthier2023} (cf. Fig.~\ref{fig:soft_overall}c inset). Such negative pulses have a negligible effect on $\bar{F}$ due to their transient and rare nature. In the subsequent sections, we will further examine the force profiles and discuss the discrepancies caused by different boundary types. The oil layer scenario, however, will be systematically analyzed separately.

	\section{Elementary impulse}
	\label{sec:element}
	Up to this point, we have drawn a correlation between transient heterogeneities and impulse bursts. This correlation suggests that the phenomenon of shear thickening is essentially a dynamically non-uniform process, similar to glass transition~\cite{Vogel2004,Xia2015}. As demonstrated in simulations~\cite{Seto2013,Henkes2016,Goyal2024,Nabizadeh2022,Naald2023,Sedes2022} and suggested by experiments~\cite{Lin2015,Lin2016}, frictional contacts form between particles under shear, subsequently developing into load bearing clusters. Recent experiments suggests these clusters cause intensive  stress response~\cite{Rathee2017,Rathee2020}  and contribute to the local increase in viscosity, thereby promoting density heterogeneities~\cite{Shi2023,Shi2024}. 
	The percolation of these frictional clusters leads to a substantial increase in the normal force response, $F(t)$. The interplay of normal stress confinement and shear profile induces reconfiguration or collapse of clusters, which subsequently results in a decrease in $F$. We interpret the fluctuations in $F(t)$ as a sequence of this microscopic process, appearing as impulse bursts. It is evident that $\bar{F}$ and $\tau_c$ in Fig.~\ref{fig:rigid_overall} and \ref{fig:soft_overall} do not suffice to capture the complexity of $F(t)$. Therefore, we introduce another time scale, the average waiting time between bursts, or equivalently, the inverse of the occurrence rate, denoted as $R$. We then define the average intensity of impulse bursts as the \textit{elementary impulse}, denoted as $Q$, which is calculated as $Q=\bar{F}/R$.

	We assume that elementary impulses occur randomly and independently in time. For a given time interval $\Delta t$, the impulse $I=\int_{t}^{t+\Dt} F(t) \mathrm{d}t$ follows a Poisson distribution. The probability is represented as $P(I/Q,\Dt)$. Here, $I/Q$ stands for the number of impulse `events', with the elementary impulse, $Q$, unknown. The waiting time between bursts then follows the corresponding exponential distribution. The average of this waiting time, $1/R$, is given by 
	\begin{equation}
		P(I=0, \Dt) = e^{-R\Dt}.
		\label{eq.QR}
	\end{equation}
	According to Eq.~\ref{eq.QR}, the average occurrence rate of the elementary impulse, $R$, can be determined by measuring $P(I=0, \Dt)$ across various $\Dt$. Note that, although $Q$ is not explicitly present in Eq.~\ref{eq.QR}, it is relevant for the measurement of $P(I=0, \Dt)$ because the null events ($I=0$) are practically determined by $I<Q$. This scenario mirrors the problem of determining the intensity threshold for an event when measuring the waiting time~\cite{Corral2004}. The choice of threshold and its influence need justification in general. Here, we employ a recursive iteration process to evaluate the value of $Q$, without predefining a threshold. Each iteration includes three steps. First, we calculate $P(I=0,\Dt)$ for a sequence of $\Delta t$ using an estimated value of $Q$ derived from the experimental $F(t)$. Next, $R$ is obtained using Eq.~\ref{eq.QR}. Finally, a newly determined $Q=\bar{F}/R$ is compared with its initial value. The iteration process concludes when the difference falls below 10\%. We emphasize that the elementary impulse defined in this way does not presume any features other than their independence. The resulting $Q$, representing the smallest independent event, may measure the impulse of an envelope of pulses or individual pulses, reflecting the characteristic of $F(t)$ at various $\srate$ (see Fig.~\ref{fig:rigid_overall}c for examples).  Readers may refer to Appendix~\ref{app:method} for additional details and discussion of the method. 
	
\begin{figure}
	\includegraphics{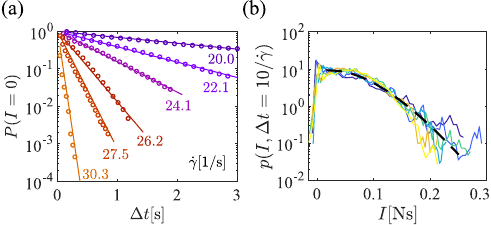}
	\caption{(a) The measured probability $P(I=0, \Dt)$ versus $ \Dt $ (circles) and the fit using Eq.~\ref{eq.QR} (solid lines) for various $\dot\gamma$. The data is collected under the rigid boundary. (b) The probability density $p(I,\Delta t=10/\srate)$ measured in experiments (solid lines) is compared to Eq.~\ref{eq.tail} with the converged $Q$ and $R$ (the dashed line) at $\srate=25\su$. The same color code as that in Fig.~\ref{fig:rigid_overall}a indicates different $h$.}
	\label{fig:dist}
\end{figure}
	
	Equation~\ref{eq.QR}, featuring the converged values of $Q$ and $R$, is plotted for various $\srate$ in Fig.~\ref{fig:dist}. This plot confirms the exponential relation presented in Eq.~\ref{eq.QR}.	The obtained $Q$ and $R$ can be further justified by comparing the tail of the measured $P(I/Q,\Dt)$ and its theoretical counterpart. For large $I$, the Poisson distribution, $P(I/Q,\Delta t)$, can be approximated with Stirling's asymptotic formula,
	\begin{equation}
		\ln P(I/Q,\Dt) \approx \frac IQ \ln\left(\frac{R\Dt Q}{I}\right) -R\Dt - \frac12 \ln\left(2\pi \frac IQ \right) + \frac IQ.
		\label{eq.tail}
	\end{equation}
	As an example, the measured distribution $P(I,\Dt=10/\srate)$ at $\srate=25\su$ is compared with Eq.~\ref{eq.tail} with $Q$ and $R$ obtained by iterating Eq.~\ref{eq.QR} in Fig.~\ref{fig:dist}. Excellent consistency is observed without free parameters.

	\section{Discussion}
	
	\begin{figure}
		\centering
		\includegraphics{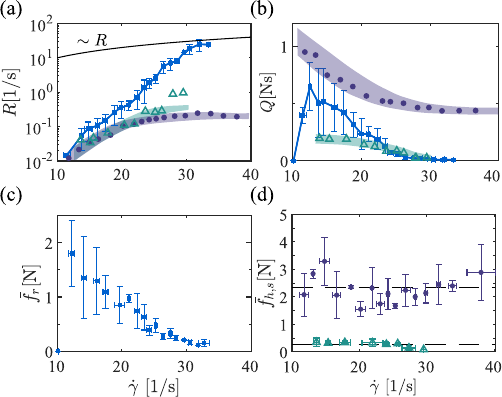}
		\caption{(a) The occurrence rate, $R$, and (b) the average intensity, $Q$, of impulses in $F(t)$ obtained through Eq.~\ref{eq.QR}. For clarity, the errors for flexible boundaries are represented by shaded areas. (c-d) The strength of individual elementary impulses, $\bar{f}=Q/\tau_c$. Diamonds $\diamond$, triangles $\triangle$ and circles $\circ$ respectively denote the rigid, hard PDMS and soft PDMS boundaries.}
		\label{fig:qr}
	\end{figure} 
	
	We present the result of $Q$ and $R$ of the elementary bursts in Fig.~\ref{fig:qr}. Henceforth, subscripts $r,\,h,\,s$ are used to denote the results for the rigid, hard PDMS and soft PDMS boundaries, respectively. The increase of $R_r$ with $\srate$ occurs at a rate significantly faster than linear, and it can be approximately described by $\sim e^{(\srate-\srate_c)/\srate^*}$, with $\srate^* = 3.5\su$. As shown in Fig.~\ref{fig:rigid_overall}c, the increase of $R_r$ at high-shear is caused by the emergence of numerous short-lived pulses with a width of $1/\srate$ (see Appendix~\ref{app:subsec}). In order to calculate $R$ in Eq.~\ref{eq.QR}, sufficient statistics around $I=0$ are necessary. The occurrence rate $R_r$ is lower than $\srate$ within the explored range (refer to Fig.~\ref{fig:qr}a), which ensures a zero background between individual pulses. It is reasonable to presume that $R$ is proportional to the boundary size, $\pi D^2/4$, just as $\bar{F}$ (refer to Appendix~\ref{app:bsize}). Hence, a larger $D$ would lead to $R>\srate$ within the range of $\srate$ under investigation. As a consequence, independent pulses would overlap in temporal occurrence, and the background of $F(t)$ would be elevated, thereby making it challenging to define the waiting time in practice. The argument provided above justifies our choice of $D$ in experiments.

	A noticeable feature of $R_r$ is its independence from the suspension height, $h$. The bursts in $F(t)$ originate from the percolation of frictional clusters. While the percolation dimension is expected to be proportional to $h$, the rate of percolation, represented by $R_r$, appears to be independent of its size. The suspension is, therefore, likely in a critical state~\cite{Henkes2016}, where frictional clusters are present at all lengths. We clarify that this does not necessarily contradict the non-power-law tail of the impulse distribution, $P(I,\Dt)$, in Eq.~\ref{eq.tail} and depicted in Fig.~\ref{fig:dist}. The normal stress probes only the percolating clusters, and $P(I,\Dt)$ represents the probability of the number of percolations occurring within $\Dt$. In contrast, the shear stress response could sample clusters of all sizes, and the corresponding dissipation displays a power-law distribution~\cite{Lootens2003}.

	Since both $\bar{F}$ and $R$ are independent on $h$, the elementary impulse $Q=\bar{F}/R$ is also scale-invariant. As shown in Fig.~\ref{fig:rigid_overall}, the relationship between $\bar{F}_r$ and $\srate$ can be approximated as linear. Meanwhile, $R_r$ depicted in Fig.~\ref{fig:qr} demonstrates an exponential growth with $\srate$. Consequently, $Q_r$ exhibits a decay rate with respect to $\srate$ that is faster than $1/\srate$. We further decompose $Q$ into two components: the duration $\tau_c$ and the strength $\bar{f}=Q/\tau_c$. Here, $\bar{f}$ represents the average force response over elementary impulses, which is distinct from $\bar{F}$, the average over the entire time series. The decrease in $Q_r$ is not determined solely by the reduction in burst duration, which transitions from an extended envelope to short pulses as illustrated in Fig.~\ref{fig:rigid_overall}. The strength, $\bar{f}_r$, decreases with $\srate$ as shown in Fig.~\ref{fig:qr}c. This observation suggests that the frequent pulses governing the fluctuations in $F_r(t)$ at high-shear are less intense than the long-lived bursts at low-shear just above $\srate_c$.
	
	For comparison, in the scenario of flexible boundaries, the rates $R_h$ and $R_s$ exhibit a power-law growth with the shear rate $\sim (\srate-\srate_c)^\alpha$, where the exponent $\alpha$ slightly greater than 1. As a consequence, the corresponding quantities $Q_h$ and $Q_s$ decay gently as $\srate$ increases. In contrast, the decrease in $Q_{h,s}$ is dominated by the shortening of $\tau_c$, leading to a roughly constant $\bar{f}_{h,s}$ as illustrated in Fig.~\ref{fig:qr}d. Further, the ratio $\bar{f}_{h}/\bar{f}_s\approx 8.2$ is comparable to the ratio of the corresponding Young's moduli of the boundaries, $E_h/E_s\approx 8.5$. As the strength $\bar{f}$ indicates the failure limit of the percolating force network, the correlation between $\bar{f}_{h,s}$ and $E_{h,s}$ suggests a unified collapse mechanism of the frictional clusters under flexible boundaries. Note that $\bar{f}_r$ falls between $\bar{f}_{h}$ and $\bar{f}_{s}$, a reminiscence of the pulse heights, $F_{high}$, shown in Fig.~\ref{fig:rigid_overall} and \ref{fig:soft_overall}. 
	
	The frictional clusters under shear have a tendency of dilatancy~\cite{Reynolds1885}. When the constant volume constraint is applied alongside a constant shear rate, a (transient) jamming event occurs. In contrast, flexible boundaries permit sufficient dilation which in turn releases the constraint and leads to unjamming. Upon a deformation on the order of $d$, the cluster collapses. Therefore, we observed a constant $\bar{f}_{h,s} \sim E_{h,s} a^{1/4} d^{3/2}$ independent on $\srate$ and proportional to $E_{h,s}$. Recall that $a$ denotes the area of the inhomogeneity. The deformability of PMDS layers promotes a significantly larger $a$ compared to what a rigid plate provides (refer to Supplemental Movie S1 and S2). This is a plausible reason for $\bar{f}_r<\bar{f}_h$ in our experiments.

	Under the rigid boundary, the unjamming/collapse of the cluster is instead realized by breaking the sliding/rotation constraints between particles in contact~\cite{Singh2020,Wang2020,Hsiao2017,Guy2018}. Such constraints can often arise due to surface asperities~\cite{Lootens2005,Hsiao2017a,Hsu2018,Schroyen2019,Pradeep2021} and adhesion~\cite{James2018,Richards2020}. Denote $\varepsilon$ as the typical scale of energy or force moment required to break individual constraints. In this scenario, the rupture of the percolating cluster assesses the hardness of interparticle interactions, $\varepsilon$, rather than the confinement stiffness for PDMS boundaries. Assuming proportionality between normal and shear stresses~\cite{Lootens2005,Brown2012}, the force moment for breaking the clusters is given by $\bar{f}_r h$, which is approximately $\sim Q_{r}\srate h$ at high-shear. A straightforward estimation suggests $\varepsilon N_{c}\sim \bar{f}_r h \sim Q_{r}\srate h$. Here, $N_{c}\sim (h/d)^{D_f}$ is the number of constraints or bonds in the cluster, and $D_f$ is the fractal dimension. Given that $\bar{f}_r$ and $Q_r$ are largely independent of $h$, this reasoning suggests the dominance of chain-like backbones, \textit{i.e.}, $D_f\approx 1$. This interpretation is consistent with the fragile state induced by shear~\cite{Cates1998}, characterized by highly anisotropic microscopic structures and force network~\cite{Foss2000,Parsi1987,Blanc2013,Seto2013}. It contrasts with isotropic aggregates~\cite{Wessel1992} that are dominated by attractive particle interactions. 
	
	The conclusion that $D_f\approx 1$ drawn above relies on the assumption of complete collapse of the cluster. In its weaker form, this assumption suggests a uniform or cooperative collapse scenario, where the number of broken bonds $n_c$ is proportional to $(h/d)^{D_f}$. Though its usage is often accepted~\cite{Gauthier2023}, this notion of uniform collapse is not always valid.	For instance, a long chain that buckles or breaks into two shorter ones can sufficiently cause percolation failure and a drop in $F(t)$. The incomplete collapse or buckling, recently suggested in experiments~\cite{Rathee2017,Rathee2020}, may consequently induce temporal correlation or strain hardening~\cite{Gadala-Maria1980,Alexander1998}, as the broken pieces promote reformation of percolating structures responding to the shear. These two scenarios of microstructure evolution could coexist and contribute differently to $\bar{f}$. While uniform collapse reveals the actual fractal structure, incomplete collapse masks it. As a consequence, the apparent $D_f$ might misrepresent the actual fractal dimension of microstructure~\cite{Nabizadeh2022}. 	Distinguishing between two collapse mechanisms requires examining features beyond the averages explored so far ($\bar{F}$ and $\bar{f}$). For instance, as  shown in Fig.~\ref{fig:dist}b, a larger $h$ systematically enhances the probability of extreme force response, likely indicating the uniform/complete collapse of the percolating structures. Detailed analysis is reserved for future studies.

	\section{Conclusion}

	We have revealed the crucial role of boundary conditions on the spontaneous inhomogeneity and the normal force response of a shear-thickening suspension under shear. The average normal force, $\bar{F}$, is raised from 0 for $\srate>\srate_c$, and its magnitude increases with the stiffness of the confinement. It is notable that the critical shear rate, $\srate_c$, is indistinguishable for both rigid and soft boundaries. This observation suggests that shear-thickening is an intrinsic characteristic of the suspension under shear. However, the consequent evolution of the underlying frictional clusters depends on the boundary flexibility. The discrepancy of boundary types also reflects on the manifestation of inhomogeneity, which is associated with the bursts in $F(t)$. With flexible boundaries, density heterogeneity spans a wider area and lasts longer, with the corresponding bursts appearing as long lifetime envelops. In contrast, for a rigid boundary, the inhomogeneity is of lower magnitude (less visible) and diminishes quickly. Moreover, as the shear rate $\srate$ increases, the bursts in $F(t)$ present as individual peaks of a width proportional to $1/\srate$. The liquid boundary is a unique case, where density heterogeneity persists with no bursts in $F(t)$. It is evident that the boundary effect on the force profile $F(t)$ exhibits complexities beyond mere averages. Thus, we decompose $F(t)$ into elementary bursts and estimate its occurrence rate, $R$. This rate is found to grow with $\srate$ faster than a linear relation, slightly for flexible boundaries and significantly under the rigid confinement. The variation of $R$ and the associated intensity of the bursts, $Q=\bar{F}/R$, indicate unique collapse mechanisms of the frictional clusters underlying the burst in $F(t)$. The dilation dominates the collapse for soft boundaries, while the contacts within the cluster are broken by shear under rigid confinements. 
	
	Our findings yield three-fold consequences. First, the normal force response and the development of inhomogeneity are primarily dictated by the microscopic structures of the particle phase and their reconfiguration. Future research may need to address the questions: of which aspect of boundary conditions should be considered for the physical process of interest. It is essential for experimental techniques like Boundary Stress Measurement~\cite{Rathee2017,Rathee2020,Hu2022,Miller2022} and capillarytron~\cite{Etcheverry2023}, where surface deformation is an inherent aspect of the measurement principle. In this context, measurement precision and the physics of interest may then present a trade-off. Second, our observation of spontaneous heterogeneity, in line with recent research~\cite{SaintMichel2018,Chacko2018,Ovarlez2020,Rahbari2021,Rathee2017,Hu2022,Gauthier2023,Shi2023,Shi2024}, necessitates a revision on mean-field theories, particularly regarding the boundary dependence. Lastly, according to our experiments with varying boundary types, the temporal fluctuation in force response can be dramatically altered. This observation suggests that certain shear boundary conditions may have the potential to maintain constant local support against external pressure, in contrast to transient impulses under steady shear between rigid plates or the violent impact scenarios~\cite{Roche2013,OyarteGalvez2017}.

	\begin{acknowledgments}
		This work is supported by NSFC (Grant number 12172277).
	\end{acknowledgments}

	\section*{Data Availability Statement}
	
	The data that support the findings of this study are available from the corresponding author upon reasonable request.

	\appendix
	
	\section{Boundary size dependence}
	\label{app:bsize}
	
\begin{figure}
		\centering
		\includegraphics{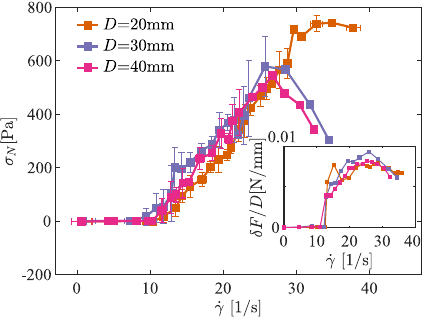}
		\caption{Force response under the rigid boundary. Pressure ${\sigma_N}$ as a function of the shear rate $\dot\gamma$ collapses for various diameters $20~\si{mm}$, $30~\si{mm}$ and $40~\si{mm}$. Data is averaged over various $h$. Inset: The ratio of the standard deviation of the force profile, $\delta F$, to the boundary diameter $D$ varies with $\dot\gamma$ for $h=3.8~\si{mm}$.}
		\label{fig:SA}
	\end{figure}

	We measured the normal force $F(t)$ for different rigid boundary areas with diameters $D$ of $20,~30,~40~\si{mm}$, respectively. The pressure ${\sigma _N} = \frac{{4\bar F}}{{\pi {D^2}}}$ is plotted versus shear rates in Fig.~\ref{fig:SA}. For all $D$, the pressure $\sigma_N$ collapses on a master curve and rises at the consistent critical shear rate $\dot\gamma_c$. The average force response $\bar{F}$ is thus proportional to the boundary area, $\pi D^2/4$, rather than the periphery. In addition, the standard deviation, $\delta F$, of $F(t)$ scales with the boundary diameter $D$, namely the square root of the area, as shown in the inset. Both scalings suggest that $F(t)$ reflects the intrinsic response of the suspension beneath the top boundary. Therefore, the edge effect can be neglected.
	
	\section{Rheology measurement}
	\label{app:rheo}
\begin{figure}
		\centering
		\includegraphics[scale=1]{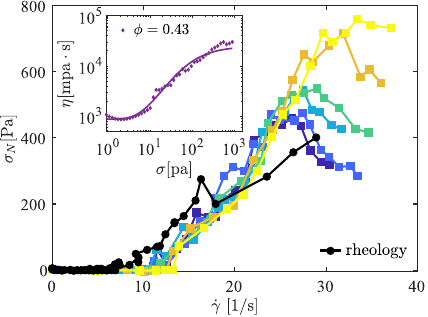}
		\caption{Normal stress $\sigma _N$ as a function of the shear rate $\dot\gamma$ for the $43\%$ volume fraction cornstarch suspension. The color lines are measurements of our orbital shaking protocol with the rigid boundary for various gap sizes, $h=$ $5~\si{mm}$, $4.7~\si{mm}$, $4.4~\si{mm}$, $4.1~\si{mm}$, $3.8~\si{mm}$, $3.5~\si{mm}$, following the same color scheme in the main text. The black points are results of the rheometer measurement. Inset: rheology curve measured in rheometer and the theoretical fit (solid lines).}
		\label{fig:SB}
	\end{figure}

	The complete dataset of average normal force response 
	$\bar{F}$ is presented in Fig.~\ref{fig:SB} for $D = 20~\si{mm}$ rigid boundary. The response diminishes beyond a certain shear rate associated with $h$. The onset shear rate for the decay decreases with $h$. We conclude that the decay of $\bar{F}$ is caused by the migration of the particle phase out of the shear zone. The decay branch is excluded from the analysis in the main text. 
	
	The normal force response is additionally confirmed with rheometry. The rheological experiment was performed using an Anton Paar MCR 302 with the parallel-plate shear cell of a gap size of $1.5~\si{mm}$. The rheological measurements were conducted in controlled shear stress mode. The onset stress of DST, $\sigma^*=\SI{9.4}{Pa}$, is obtained by fitting the viscosity-stress data with the Wyart-Cates model~\cite{Wyart2014} (Fig.~\ref{fig:SB} inset). The normal stress measured in the rheometry is plotted versus the average shear rate in Fig.~\ref{fig:SB}, which aligns with the our measurement. The rise of the normal stress is observed around $9.2\su$, which is comparable with the critical shear rate $\dot\gamma_c=11.4 \pm 1.2\su$ observed in our orbital-shear protocol. The agreement of the average response between rheometer and our protocol thus validates the control parameter $\srate$. The remaining difference may result from the non-uniform and unsteady shear rate in the rheometer configuration. In addition, the considerably smaller gap size of the rheometer may contribute to the quantitative differences as well. 
	
	\section{\label{app:subsec}Pulse width and autocorrelation time $\tau_c$}
\begin{figure}
		\centering
		\includegraphics[scale=1]{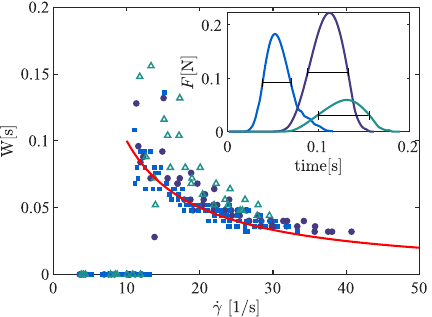}
		\caption{The full width at half peak height of individual pulses, $W$, versus shear rate $\dot\gamma$ under rigid, hard PDMS, and soft PDMS boundaries. The same symbols and colors as in Fig.~\ref{fig:rigid_overall}b and Fig.~\ref{fig:soft_overall}b are used. The red solid curve denotes ${1}/{{\dot \gamma }}$. Inset: Typical peak shape at $\dot \gamma  = 25\si{s^{ - 1}}$ for the three solid boundary types. Colors correspond to that in the main plot. }
		\label{fig:SC}
	\end{figure}

	We identify peaks in $F(t)$ via local maxima that exceed the background noise level by at least a factor of 10. The peaks of most probable height are averaged to represent the typical shape, as shown in the inset of Figure~\ref{fig:SC}. These contours represent the typical normal force shapes under different boundaries, called the typical normal force pulses. Under rigid boundaries, pulses arise instantaneously, while the decay is relatively gradual. For the PDMS boundary, both the rise and decay are slow. 	
	The full width at half peak height, $W$, characterizes the duration of individual pulses. We find $W\approx \frac{1}{{\dot\gamma}}$ for all solid boundaries. 
	
\begin{figure}
		\centering
		\includegraphics{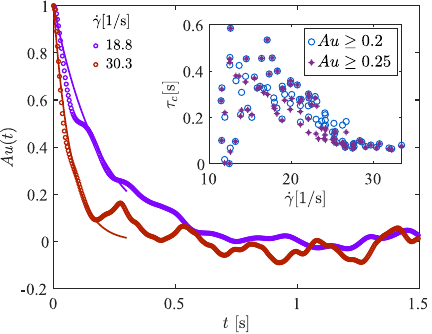}
		\caption{The autocorrelation $Au(\tau)$ computed for two shear rates. The solid lines represent the exponential decay fits in the short-time region. Inset: The extracted $\tau_c$ from $Au(\tau)$ for two definitions of short-time region.}
		\label{fig:autocorr}
	\end{figure}

	In the main text, we use autocorrelation time $\tau_c$ to measure the duration of impulses. For a time series of length $T$, the autocorrelation is computed by
	\[Au(t) = \frac{1}{T\delta F^2}\sum_{\tau=1}^{T-t} \left(F(\tau)-\bar{F}\right)\left(F(\tau+t)-\bar{F}\right).\]
	Below $\srate_c$, $Au(t)$ is effectively zero for the white noise $F(t)$. Beyond $\srate_c$, $Au(t)$ decays from 1 monotonically in the short-time regime. Here, we define the short-time regime according to $1>Au(t)>0.2$ and fit it with $e^{t/\tau_c}$ to obtain $\tau_c$. This choice of short-time region is empirical, which in general covers the rapid decay of $Au(t)$, thus sampling the individual bursts. Increasing the lower bound of the region from $Au=0.2 $ to 0.25 results in only subtle changes in $\tau_c$ (see Figure.~\ref{fig:autocorr}).
	
	Note that $\tau_c$ is always larger than $W$, since it samples relatively long envelopes (groups of pulses) in $F(t)$ as well. This is particularly evident under rigid boundaries. When the pulses exhibit little correlation at high-shear, $\tau_c$ becomes proportional to $W\approx\frac{1}{{\dot \gamma }}$, as shown in Fig.~\ref{fig:rigid_overall}b.

	\section{Calculation of $R$ and $Q$}
	\label{app:method}
	The occurrence rate $R$ can be calculated in Eq.~\ref{eq.QR}, if the probability of no impulse events, $P_0(\Dt) = P(k=0,\Dt)$, is measured. Here, $k$ represents the number of impulse events during $\Dt$. In practice, an event threshold must be introduced, such that  $I<Q$ defines the null events. The elementary impulse $Q$ is thus essential for the calculation. In addition to Eq.~\ref{eq.QR}, $Q$ and $R$ are linked by the average force: $\bar{F} = QR$. Therefore, we recursively determine $R$ and $Q$, as described in the main text. The initial value of $Q$ is chosen as $\bar{F}\tau_c$, where $\bar{F}$ and $\tau_c$ are the average of $F(t)$ and the autocorrelation time respectively.
	
	By rewriting Eq.~\ref{eq.QR}, the value of $Q$ satisfies 
	\begin{equation}
		\frac{\bar{F}}{Q}\Dt = R\Dt =  - \ln(P_0(\Dt)).
		\label{eq.recQ}
	\end{equation}
	We justify the convergence of Eq.~\ref{eq.recQ} as follows. With sufficient statistics around $I\approx 0$ (cf. Fig.~\ref{fig:dist}b), we approximate $P_0(\Dt)$ as proportional to $Q$, \textit{i.e.}, $P_0(\Dt)=p_0(\Dt) Q$. Substituting this approximation into Eq.~\ref{eq.recQ}, the converged $Q$ satisfies
	\begin{equation}
		\frac{\bar{F}}{Q}\Dt =  - \ln\left( p_0(\Dt) Q \right)
		\label{eq.map}
	\end{equation}
	Given experimental parameters and $\Dt$, all quantities except $Q$ are constant. For convenience, we define $x=1/Q$. The iteration scheme is then described by
	\begin{equation}
		\bar{F}\Dt\; x_{n+1} = \ln\left( \frac{x_n}{p_0(\Dt) } \right)
		\label{eq.itsch}
	\end{equation} 
	
	The number of roots for Eq.~\ref{eq.map}, which is the number of intersections between a linear and logarithm-like function, depends on $\Delta t$. As $\Delta t$ is varied, without loss of generality, we consider the cases where there are either no roots or two. In the former case, $p_0>1/(e\bar{F}\Dt)$. Recall that $P_0$ represents the probability of null impulse events, and $\bar{F}\Dt$ gives the average impulse of the distribution. Therefore, the no-solution case typically suggests a notable probability around zero along with a large average. This scenario likely occurs for distributions significantly deviating from the monomial shape exhibiting peaks around $I=0$ and some finite value of $I$. However, we have not observed such shapes for the range of $\Dt$. In the latter case, we denote the two roots as $Q_1 < Q_2$. It is readily to show that $Q_1$ is a stable fixed point that can be recovered by the iteration scheme in Eq.~\ref{eq.itsch}.
	
	The elementary impulses identified by our method represent the smallest independent bursts in $F(t)$. One advantage of this approach is insensitivity to noise levels, as background noise largely cancels out in the calculation of $I(\Delta t)$. This allows us to process the data with relatively high noise levels under the soft PDMS boundaries, where thresholding is difficult. When the noise levels are too high, the occurrence rate $R$ could still be overestimated, causing the high shear results for soft PDMS in Fig.~\ref{fig:qr}a to appear as outliers.
	
	\bibliography{boundary_constraints}
	
\end{document}